\documentclass[aps,prl,showpacs,twocolumn, superscriptaddress]{revtex4-1}

\usepackage{graphicx}
\usepackage{dcolumn}
\usepackage{bm}
\usepackage{color}
\usepackage{colortbl}
\usepackage{amsmath}
\usepackage{amssymb}
\usepackage{mathrsfs}  

\usepackage{multirow}

\newcommand{\tabincell}[2]{\begin{tabular}{@{}#1@{}}#2\end{tabular}}

\begin{document}

\title{Extended Wiener-Khinchin theorem for quantum spectral analysis}
\author{Rui-Bo Jin}
\affiliation{Laboratory of Optical Information Technology, Wuhan Institute of Technology, Wuhan 430205, China\\}
%
\author{Ryosuke Shimizu}
\email{Corresponding author: r-simizu@uec.ac.jp}
\affiliation{The University of Electro-Communications, 1-5-1 Chofugaoka, Chofu, Tokyo, Japan}

\date{\today }

\begin{abstract}
The classical Wiener-Khinchin theorem (WKT), which can extract  spectral information by classical interferometers through Fourier transform, is a fundamental theorem used in many disciplines.
However, there is still need for a quantum version of WKT, which could connect correlated biphoton spectral information by quantum interferometers. Here, we extend the classical WKT  to its quantum counterpart, i.e., extended WKT (e-WKT), which is based on two-photon quantum interferometry.
According to the e-WKT, the difference-frequency distribution of the biphoton wavefunctions can be extracted  by applying a Fourier transform on the time-domain Hong-Ou-Mandel interference (HOMI) patterns, while the  sum-frequency distribution can be extracted by applying a Fourier transform on the time-domain NOON state interference (NOONI) patterns.
We also experimentally  verified the WKT and e-WKT in a Mach-Zehnder interference (MZI), a HOMI and a NOONI.
This theorem can be directly applied to quantum spectroscopy, where the spectral correlation information of biphotons can be obtained from time-domain quantum interferences by Fourier transform. This may open a new pathway for the study of light-matter interaction at the single photon level.
\end{abstract}

\pacs{42.65.Lm, 03.65.Ud, 42.50.St, 42.50.Dv }


\maketitle

\textbf{\emph{Introduction}}
The Wiener-Khinchin theorem (WKT), which expresses the power
spectrum in terms of the autocorrelation function by Fourier transformation, was proved by N. Wiener \cite{Wiener1930}  and by A. Khinchine \cite{Khintchine1934}  in the 1930s.
The WKT is a fundamental theorem used in many disciplines, such as statistics, signal analysis and optics, etc.
Especially in modern optics, thanks to the Wiener-Khinchin theorem, the interferometric spectrometer technology (also called Fourier transform spectrometry) has been well established \cite{Davis2001}. For example, it is possible to extract the spectral information of  light by making a Fourier transform on its time-domain  Mach-Zehnder  interference (MZI) or Michelson interference (MI) patterns.
Such interferometric spectrometers are especially useful for simultaneously collecting high spectral resolution data over a wide spectral range. This provides a significant advantage over a dispersive spectrometer which measures intensity over a narrow range of wavelengths at a time.
The Fourier Transform Infrared Spectroscopy (FTIR) has been commercially used in chemical analysis, polymer testing and pharmaceutical analysis, etc. \cite{Griffiths2007}.

With the development of quantum optics in the last several decades, several new interferometries have been demonstrated, such as the  Hong-Ou-Mandel  interference (HOMI) \cite{Hong1987}  and NOON state interference (NOONI) \cite{Boto2000} using biphotons from spontaneous parametric down conversion (SPDC).
The HOMI has been widely used in  quantum optical coherence tomography \cite{Nasr2003}, dispersion cancellation \cite{Franson1992, Steinberg1992}, tests of the  indistinguishability of  two incoming photons \cite{Santori2002, Beugnon2006, Mosley2008a, Mosley2008b, Jin2011, Ansari2014}, measurement of the biphoton wave function  \cite{Chen2015}, frequency conversion \cite{Kobayashi2016}, and discrete frequency modes generation \cite{Jin2016QST}.
The NOONI has been widely used in quantum lithography \cite{Boto2000, Edamatsu2002}, quantum high-precision measurement \cite{Giovannetti2004}, quantum microscopy \cite{Ono2013, Israel2014, Jin2016SR},   and  error correction \cite{Bergmann2015}.
These two kinds of biphoton interferometries are totally quantum effect \cite{Giovannetti2002}, which is different from the classical one-photon MZI patterns.
This naturally gives rise to the question: Is it possible to construct a quantum interferometric spectrometer based on the quantum interference patterns?
In other words, what kind of spectral information can be extracted from the time-domain biphoton HOMI and NOONI patterns?

\begin{figure*}[tbp]
\centering
\includegraphics[width= 0.9\textwidth]{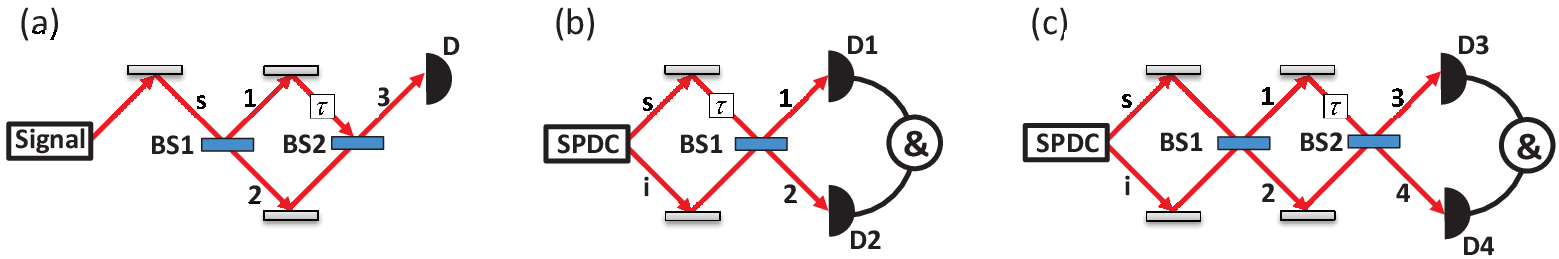}
\caption{  Model of the experimental setup. (a) Mach-Zehnder interferometer (MZI), (b) Hong-Ou-Mandel interferometer (HOMI), (c) NOON-state interferometer (NOONI). $s$ and $i$ indicate the signal and idler photons. Both the signal and idler photons from SPDC are used for HOMI and NOONI, while only the signal photons  are used for MZI.
}
\label{model}
\end{figure*}

To answer this question in this work, we first provide a multi-mode theory  for the MZI, HOMI and NOONI, with the model shown in Fig.\,\ref{model}(a-c).
Then, we expand the classical WKT based on MZI into  an extended WKT (e-WKT)  based on HOMI and NOONI.
Using this e-WKT, it is possible to extract the difference-  or  sum-frequency information between the constituent photons from the time-domain HOMI and NOONI patterns.
Finally, we verified our theory experimentally by measuring the  MZI/HOMI/NOONI  patterns and two-photon spectral intensity distribution.

\textbf{\emph{Theory}}
In this Letter,  we expanded the traditional  WKT to its quantum version.
First, let us consider the classical WKT in the scenario of an MZI as shown in Fig.\,\ref{model}(a).
%
%
%
As calculated in the \textbf{Supplementary Information}, in an MZI, the one-photon detection probability is determined by
\begin{equation}\label{eqs:P1}
P_1(\tau ) = \frac{1}{2}[1 + \int_{-\infty}^\infty  {d\omega } \left| {f_1(\omega )} \right|^2 \cos(\omega \tau )],
\end{equation}
where $f_1(\omega )$ is the one-photon spectral amplitude.
The conventional WKT can be written in the form of Fourier transform:
\begin{equation}\label{eqs:WKT}
F_1(\omega) \equiv \left| {f_1(\omega )} \right|^2 = \frac{1}{2\pi}\int_{-\infty}^\infty  {d\tau } G_1 (\tau )e^{i\omega \tau },
\end{equation}
where  $F_1(\omega) \equiv \left| {f_1(\omega )} \right|^2$ is the one-photon spectral intensity,
and $G_1 (\tau ) \equiv \int_{-\infty}^\infty  {d\omega } \left| {f_1(\omega )} \right|^2 e^{ - i\omega \tau }$   denotes the first-order correlation function.
Here we adopt the definition of $  P(\tau)= \frac{1}{2}[1 + Re \{ G_1(\tau) \}]$, which is identical to that in Ref. \cite{Loudon2000}.
Based on this WKT, we can extract the  frequency information of the photon source from the time-domain MZ interference pattern.

Next, we consider the quantum counterpart of WKT, i.e., the e-WKT, which is based on HOMI shown in Fig.\,\ref{model}(b) and NOONI shown in Fig.\,\ref{model}(c).
As calculated in the \textbf{Supplementary Information}, the two-photon detection probability $P_2^{\pm}(\tau)$ is
\begin{equation}\label{eqs:P2}
\begin{array}{lll}
 P_2^{\pm}(\tau)
  &=& \frac{1}{2}[1 \pm \int_{-\infty}^\infty \int_{-\infty}^\infty  {d\omega _s } d\omega _i {\rm{|}} f_2(\omega _s ,\omega _i ) {\rm{|}}^2 \\
  && \cos[(\omega _s  \pm \omega _i )\tau]], \\
 \end{array}
\end{equation}
where  $f_2(\omega _s ,\omega _i )$ is the two-photon spectral amplitude for the signal photon with a frequency of $\omega _s$ and idler photon with a frequency of $\omega _i$. $P_2^{+}$ is for NOONI, while  $P_2^{-}$ is for HOMI.
The e-WKT can also be written in the form of a Fourier transform:
\begin{equation}\label{eqs:e-WKT}
F_2^{\pm}(\omega _{\pm}) = \frac{1}{2\pi} \int_{-\infty}^\infty d\tau G_2^{\pm}(\tau) e^{i \omega _{\pm} \tau}, \\
\end{equation}
where $\omega _{\pm}= \omega _s  \pm \omega _i $ and $F_2^{\pm}(\omega _ \pm) \equiv \frac{1}{2}\int_{-\infty}^\infty d\omega _\mp | f_2(\omega _s ,\omega _i ) |^2 $ is the sum- or difference-frequency spectrum intensity of the two-photon state, i.e., the projection of $| f_2(\omega _s ,\omega _i ) |^2$ onto the diagonal or anti-diagonal axis. $G_2^{\pm}(\tau)  \equiv  \int_{-\infty}^\infty d\omega_{\pm} F_2(\omega _{\pm})  e^{-i \omega _{\pm} \tau}$ is the second-order correlation function. $G_2^{+}$ is for NOONI, while  $G_2^{-}$ is for HOMI. $ P_{2}^{\pm}(\tau)= \frac{1}{2}[1 \pm Re \{G_2^{\pm}(\tau)\}]$.
Based on  this e-WKT, we can obtain the sum- (differential) frequency information of the two-photon source from the time-domain NOONI (HOMI) pattern.

\textbf{\emph{Experiment and Results}}
Next, we experimentally compare the  e-WKT  in Eq.(\ref{eqs:e-WKT}) with the WKT in Eq.(\ref{eqs:WKT}).
First, we carry out three types of interference experiments, i.e., MZI, HOMI and NOONI in the time domain; and we perform Fourier transformation on the time domain data so as to obtain the spectral information, especially the spectral bandwidths.
Secondly, we measure the two-photon spectral intensity (TSI) distribution of our biphotons from SPDC; and we project the TSI data onto the x-axis, the diagonal axis  and the anti-diagonal axis, respectively, so as to obtain the spectral bandwidth on each axis.
Finally,  we verify the e-WKT and WKT  by comparing the experimentally measured spectral bandwidths and those calculated using  e-WKT or WKT.

%
\begin{figure*}[tbp]
\centering
\includegraphics[width= 0.99\textwidth]{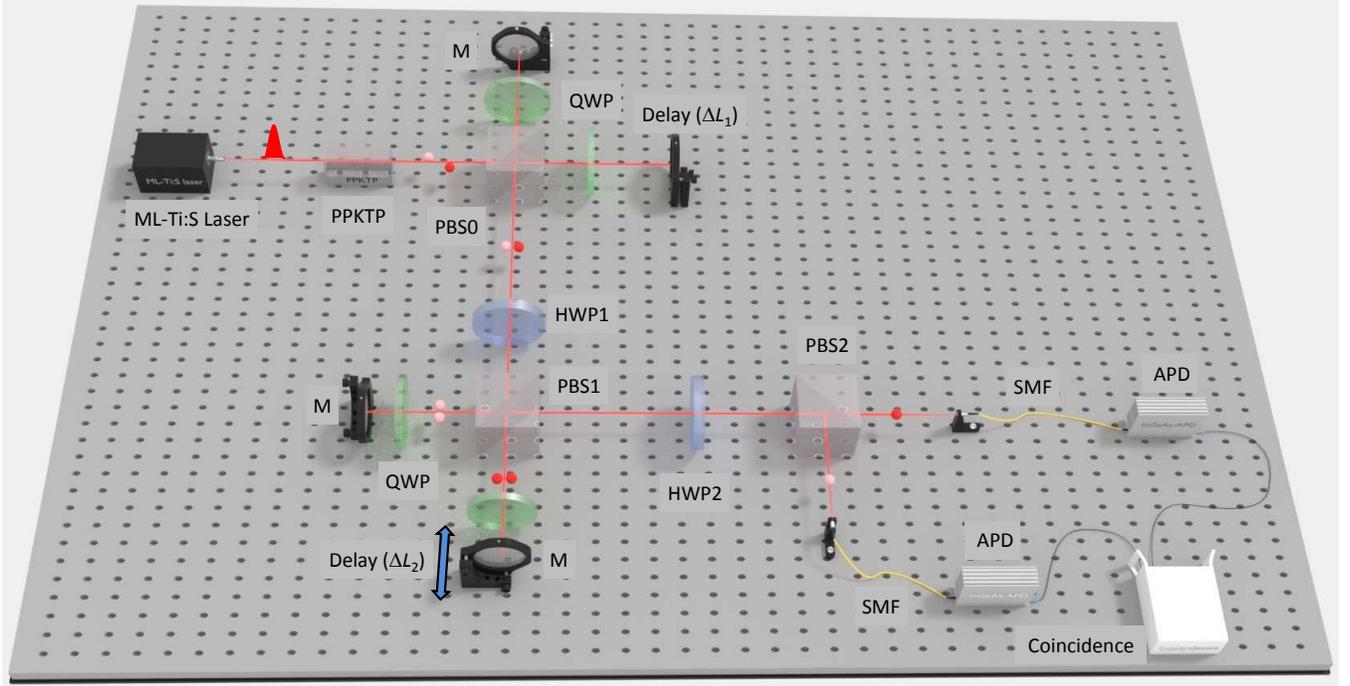}
\caption{  The experimental setup. M = mirror, QWP = quarter wave plate, HWP = half wave plate, PBS = polarization beam splitter, SMF = single-mode fiber, APD = avalanche photodiode.
}
\label{setup}
\end{figure*}
%

The setups for measuring the MZI, HOMI and NOONI are shown in  Fig.\,\ref{setup}, and are similar to the setups reported in previous studies \cite{Shimizu2009, Bisht2015}. %
Pulses of 120-fs in length at 792 nm  are used to pump a 30-mm-long PPKTP crystal for a type-II collinear SPDC.
The PPKTP crystal can satisfy the group-velocity-matching (GVM) condition at telecom wavelength \cite{Jin2013OE, Konig2004, Evans2010, Gerrits2011, Eckstein2011, Bruno2014}.
Thanks to the GVM condition,  we can manipulate two-photon spectral distributions and generate biphotons with positive spectral correlation. In practice, the FWHM of $F_2^+$ is determined by the pump laser spectrum while that of $F_2^-$ is determined by the crystal length.
The signal and idler photons generated from SPDC have the degenerate wavelengths and orthogonal polarizations.
To compensate for their different group velocities due to the birefringence of the nonlinear crystal, the downconverted biphotons pass  through a timing compensator which is composed of a polarization beam splitter (PBS0), two quarter wave plates (QWP, at 45\,$^{\circ}$) and two mirrors.
One of the mirrors is set on a stepping motor to prepare an optical path delay of $\Delta L_1$.
Then, the polarizations of biphotons are mixed at a half wave plate (HWP1, at 0\,$^{\circ}$ for HOMI, or at 22.5\,$^{\circ}$ for NOONI) before they are input into a Michelson interferometer that has the same configuration as the timing compensator.
After that, the polarizations of biphotons are mixed again at HWP2 (fixed at 22.5\,$^{\circ}$) and separated at PBS2.
Finally all the photons are coupled into two single-mode fibers (SMF) and detected by two InGaAs avalanche photodiodes (APDs), which are connected to a coincidence counter.
This setup is versatile: by keeping HWP1 at 0\,$^{\circ}$, the setup is for HOMI; by rotating HWP1 to 22.5\,$^{\circ}$, the setup can measure NOONI; by blocking one arm of the delay line ($\Delta L_1$), the setup is ready for a one-photon MZI. So, this setup can realize all the models in Fig.\,\ref{model}.

%
\begin{figure*}[tbp]
\centering
\includegraphics[width= 0.95\textwidth]{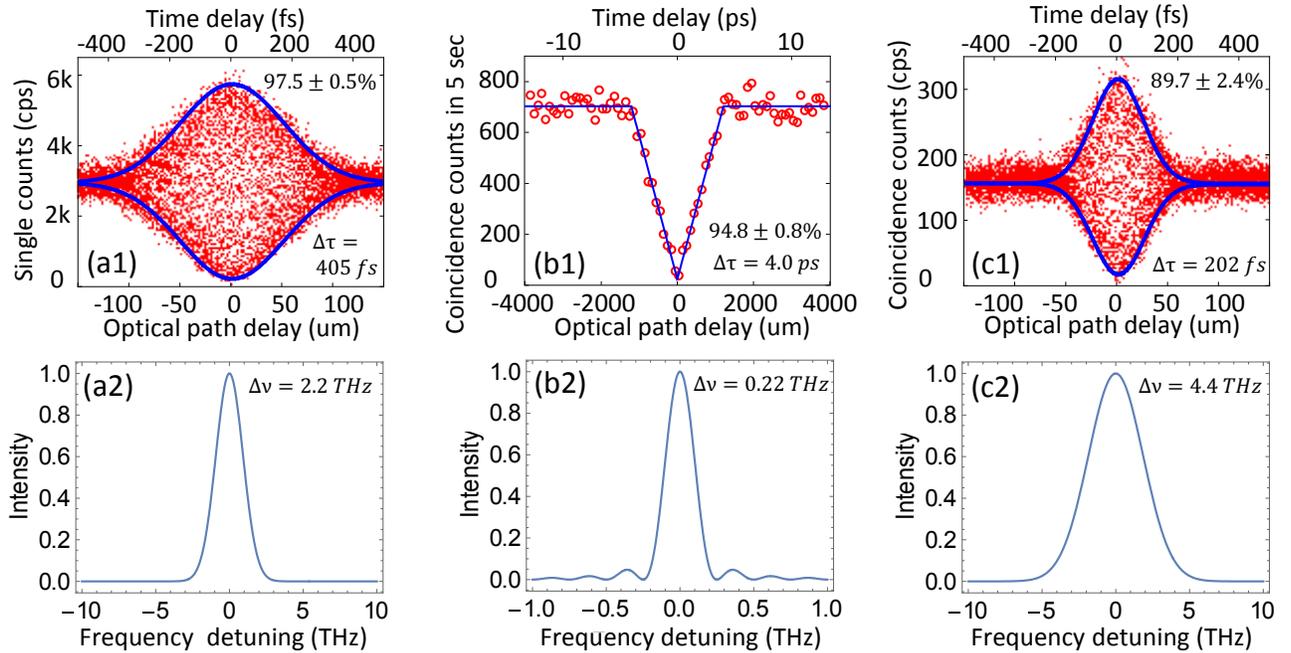}
\caption{  The time domain interference patterns and their Fourier transformed frequency distribution.   The first row shows the  experimentally measured interference patterns: (a1) Mach-Zehnder interference pattern, (b1) Hong-Ou-Mandel interference pattern, (c1) NOON state interference pattern. The visibility and temporal FWHM ($\Delta \tau$) are shown in each figure. The figures in the second row (a2, b2, c2) show the corresponding frequency distribution, calculated from  interference patterns (a1, b1, c1) by the Fourier transformation. The spectral FWHM ($\Delta \nu$) is shown in each figure.
}
\label{pattern}
\end{figure*}
%
The measured  interference patterns are shown in  Fig.\,\ref{pattern}(a1-c1).  The MZI pattern in Fig.\,\ref{pattern}(a1) is fitted by a Gaussian function with a full-width-at-half maximum (FWHM) of 405 fs and visibility of 97.5 $\pm$ 0.5 \% for the upper and lower envelopes.  The HOMI in Fig.\,\ref{pattern}(b1) has a triangle profile with an FWHM of 4 ps and visibility of 94.8 $\pm$ 0.8\%. The NOONI in  Fig.\,\ref{pattern}(c1) is fitted by a triangle function with an FWHM of 202 fs and visibility of 89.7 $\pm$ 2.4\%. The uncertainties for the visibility were added by assuming Poissonian statistics of the coincidence counts.
Although we can estimate a center frequency of the spectral peak from a fringe period, it is hard to determine the spectral peak position with high accuracy due to the instability of the interferometers over the long accumulation time in the photon counting measurements.
Thus, here we focus on extracting the spectral shape, and just adopt the envelop shape of the interference patterns.

Figure\,\ref{pattern}(a2-c2) shows the corresponding frequency distribution, which is  calculated from  the interference patterns by the Fourier transformation.   Figure\,\ref{pattern}(a2) shows the corresponding  spectral information of Fig.\,\ref{pattern}(a1), with an FWHM of 2.2 THz in frequency.   Figure\,\ref{pattern}(b2) has a $sinc^2$ profile with an FWHM of 0.22 THz in frequency, which is determined by its Fourier transform pair, i.e., the triangle-profile data in Fig.\,\ref{pattern}(b1).     Figure\,\ref{pattern}(c2) also has a Gaussian distribution with an FWHM of 4.4 THz.

\begin{figure*}[tbp]
\centering
\includegraphics[width= 0.95\textwidth]{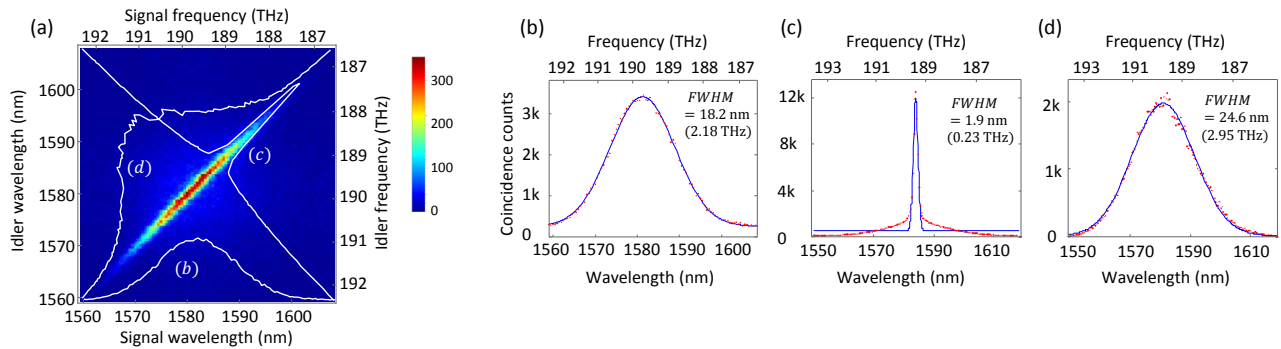}
\caption{  The experimental TSI and its projections onto three axes. (a) The experimentally measured two-photon spectral intensity (TSI) of the signal and idler photons from SPDC.  The projections of the TSI onto the x-axis (b),  anti-diagonal axis (c) and diagonal axis (d) are shown. The corresponding FWHM values are 18.2 nm (2.18 THz), 1.9 nm (0.23 THz) and 24.6 nm (2.95 THz), respectively.
}
\label{tsi}
\end{figure*}

Secondly, we measured the TSI in an experiment using the same setup as reported in previous studies \cite{Shimizu2009, Jin2013OE}.
The TSI is measured by using two center-wavelength-tunable bandpass filters (BPF), which have a filter function of Gaussian shape with an FWHM of 0.56 nm and a tunable central wavelength from 1560 nm to 1620 nm \cite{Shimizu2009, Jin2013OE, Bisht2015}.
The two single photon detectors used in this measurement are two InGaAs avalanche photodiode (APD) detectors (ID210, idQuantique), which have  a quantum efficiency of around 20\% with a dark count around 2 kHz.
To measure the TSI of the photon pairs, we scanned the central wavelength of the two BPFs, and recorded the coincidence counts. The two BPFs were moved  0.1 nm per step and 60 by 60 steps in all. The coincidence counts were accumulated for 5 seconds for each point.
The measured TSI is shown in  Fig.\,\ref{tsi}(a), and was obtained by scanning two center-wavelength-tunable bandpass filters. The projected spectral distribution onto the x-axis, anti-diagonal direction and diagonal-direction are labeled in Fig.\,\ref{tsi}(a) and in Figs.\,\ref{tsi}(b-d), respectively. The corresponding FWHM values are 18.2 nm (2.18 THz),  1.9 nm (0.23 THz) and 24.6 nm (2.95 THz), respectively.
%
%
%
%
%
%

Finally, we compared the spectrally measured FWHM values from the TSI data with the  FWHM values calculated using e-WKT or WKT in Table\,\ref{table1}.
The first row in Table\,\ref{table1} shows the $\Delta t$, which is the FWHM of the MZI/HOMI/NOONI patterns in Figs.\,\ref{pattern}(a1-c1).
The second row shows the corresponding frequency bandwidth in Figs.\,\ref{pattern}(a2-c2), as calculated from  interference patterns by Fourier transformation.
The third row shows the FWHM of the projection distributions in Figs.\,\ref{tsi}(b-d).

The classical WKT is well verified  using the data in the first column in  Table\,\ref{table1}, since the 2.2 THz bandwidth from the MZI data corresponds well with the 2.18 THz bandwidth from the TSI data.
The e-WKT values for differential frequency distribution are also well verified using data in the second column. The 0.22 THz bandwidth from the HOMI data corresponds well with the 2.23 THz bandwidth from the TSI data, proving the validity of our theory.
The data in the third column also partially verified the  e-WKT for the sum frequency distribution, since the 4.4 THz bandwidth from the NOONI data is a little bigger than the 2.95 THz using TSI data. This may have been due to the fact that the InGaAs APDs have a large dark count (around 2 kHz), a low detection efficiency (around 20\%) and  a strong wavelength dependency of the detection efficiency around 1600 nm, and as a result, the large background counts decreased the FWHM  along the diagonal-direction in the TSI measurement.
The sum frequency bandwidth of 4.4 THz, obtained from e-WKT, is in good agreement with the theoretically expected value. This means that measurement through e-WKT provides accurate spectral information while the direct spectral measurements may suffer from the  detector characteristics.

\begin{table*}[tbp]
\begin{tabular}{c|c|c|c}
\hline\hline
\tabincell{c}{Time domain \\ interference data \\ FWHM}  & \tabincell{c}{\textbf{MZI} \\  $\Delta t = 405$ fs}& \tabincell{c}{\textbf{HOMI} \\  $\Delta t = 4.0$ ps}& \tabincell{c}{\textbf{NOONI} \\  $\Delta t = 202$ fs} \\
\hline\hline
\tabincell{c}{ Expected \\ frequency width \\using (e-)WKT }
& \tabincell{c}{ \textbf{MZI } \\   $\Delta \nu  = 2.2$ THz}
 & \tabincell{c}{\textbf{HOMI}  \\  $\Delta \nu  = 0.22$ THz }
& \tabincell{c}{\textbf{NOONI} \\  $\Delta \nu  = 4.4$ THz}  \\
\hline\hline
\tabincell{c}{ Frequency domain \\TSI data \\ FWHM}
& \tabincell{c}{ \textbf{Project on x axis } \\  $ \Delta \lambda  = 18.2$ nm \\ $\Delta \nu  = 2.18$ THz}                & \tabincell{c}{\textbf{Project on anti-diagonal}  \\ $ \Delta \lambda  = 1.9$ nm \\ $\Delta \nu  = 0.23$ THz }
& \tabincell{c}{\textbf{Project on diagonal } \\ $\Delta \lambda  = 24.6$ nm \\ $\Delta \nu  = 2.95$ THz}  \\
\hline\hline
\end{tabular}
\caption{\label{table1}
Comparison of the time domain data and spectral domain data. The parameters in the first row are from Figs.\,\ref{pattern}(a1-c1). The parameters in the second row are from Figs.\,\ref{pattern}(a2-c2). The parameters in the third row are from Figs.\,\ref{tsi}(b-d).  }
\end{table*}

\textbf{\emph{Discussion}}
The e-WKT expressed in Eq.(\ref{eqs:e-WKT})  and the traditional WKT (in Eqs.(\ref{eqs:WKT})) are unified in  form.
Both the WKT and e-WKT correspond to one-dimensional Fourier transform, which builds a bridge between the spectral distribution in intensity and time-domain  interference patterns.
However, the WKT deals with uncorrelated photons, while the e-WKT is used with correlated biphotons.
In the e-WKT,  the TSI (in intensity, not amplitude) is directly related to the time-domain interference patterns. This feature is of great importance, because there is no need to measure the amplitude information, which is usually phase-sensitive and difficult to measure experimentally.

It should be noted that, in the deduction of the e-WKT in Eq.(\ref{eqs:e-WKT}), we assumed the SPDC source had a symmetric distribution, i.e.,  $f(\omega _s, \omega _i)=f(\omega _i, \omega _s)$. Under this condition, the e-WKT has a quite simple and elegant form. If this condition is not satisfied, however, Eq.(\ref{eqs:e-WKT}) will have a more complex form, which we will address in a future work.

We can now answer the question posed in the introduction: It is indeed possible to realize a quantum interferometric spectroscopy that can extract  difference- or sum-frequency information between two photons  from the time-domain HOMI and NOONI patterns.
Based on the classical WKT, it is possible to reconstruct the  spectral information of optical pulses by doing MZI.
In another words, we built a classical interferometric spectroscopy technology based on WKT.
Base in turn on this e-WKT, it is possible to establish a quantum interferometric spectroscopy technology, and many promising applications become possible.
One immediate application of the e-WKT is for nonlinear spectroscopy at the single photon level, such as for entangled photon generation using an excitonic system \cite{Edamatsu2004}. Although exciton physics has been well-studied by classical spectroscopy, a spectral entanglement of photons may contain rich information on excitonic properties, which never extract by classical spectroscopy, and allow us to discuss a new type of light-matter interaction. The Fourier transform spectroscopy based on the e-WKT is expected to be a powerful tool for investigating nonlinear light-matter interactions at the single photon level. In the future, we may apply this technique not only for biphotons from condensed matter but also for faint emissions from biological samples.

\textbf{\emph{Conclusion}}
We theoretically and experimentally demonstrated an extended Wiener-Khinchin theorem (e-WKT) .
Unlike the classical WKT, which can bridge the time-domain autocorrelation function and frequency-domain spectral intensity by Fourier transform for the classical uncorrelated photons, this new theorem can establish such a bridge for the quantum correlated biphotons:  the sum- or difference-frequency information between the constituent photons can be extracted from the time-domain HOMI or NOONI patterns.
This theorem can be directly applied to quantum spectroscopy in which the spectral correlation information of biphotons can be obtained from time-domain quantum interference by Fourier transform.

\textbf{\emph{Acknowledgements}}
We thank Zhen-Yu Wang and Wenxian Zhang for helpful discussions.
R.S. acknowledges support from the Research Foundation for Opto-Science and Technology, Hamamatsu, Japan. R.J. is supported by a fund from the Educational Department of Hubei Province, China (Grant No. D20161504).

%

\onecolumngrid
\clearpage

\subsection*{Supplementary Information to \\  Extended Wiener-Khinchin theorem for quantum spectral
analysis}

\subsection*{S1: The conventional Wiener-Khinchin theorem based on Multi-mode Mach-Zehnder interference}
\begin{figure}[bp]
\includegraphics[width=0.95 \textwidth]{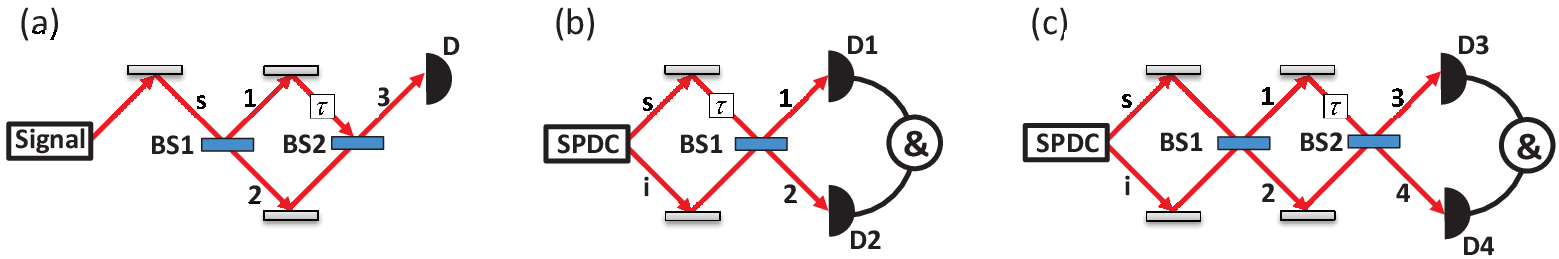}
\caption{The setups. (a)Mach-Zehnder (MZ) interference, (b) Hong-Ou-Mandel (HOM) interference, (c) NOON-state interference.}
\label{s1}
\end{figure}

In Section S1,  we provide a multi-mode theory for Mach-Zehnder (MZ) interference.
Based on the equations of this interferometry, we can construct the conventional Wiener-Khinchin theorem (WKT), which is the foundation for the classical interferometric spectroscopy.
The setup of the MZ interference is shown in Fig.\,\ref{s1}(a).
Assume there is a single photon state  $\left| \psi  \right\rangle$, which has a frequency distribution (i.e., one-photon spectral amplitude) of $f(\omega _s )$
\begin{equation}\label{eq0}
\left| \psi  \right\rangle  = \int_0^\infty  {d\omega _s } f(\omega _s )\hat a_s^\dag  (\omega _s )\left| 0 \right\rangle.
\end{equation}
where $\hat a_s^\dag$ is the creation operator and $\omega _s$ is the angular frequency.

The photons from the single photon source  are split by the first 50/50 beam splitter (BS1) and then pass through path 1 and 2. Then, after an optical delay $\tau$, the photons combine at the second 50/50 beam splitter (BS2). The photons at the output port 3 of BS2 are detected by a single photon detector D.
The detection field operator of detector (D) is $\hat E^{( + )} (t) = \frac{1}{{\sqrt {2\pi } }}\int_0^\infty  {d\omega } \hat a(\omega )e^{ - i\omega t}$, where $\hat a(\omega )$ is the annihilation operator for the frequency $\omega $ in the detection filed.
By considering the relation of $\hat a(\omega ) = \frac{1}{{\sqrt 2 }}[\hat a_1 (\omega )e^{ - i\omega \tau }  + \hat a_2 (\omega )] = \frac{1}{2}\hat a_s (\omega )(e^{ - i\omega \tau }  + 1)$, where $\hat a_1$ and $\hat a_2$   are the annihilation operators for path 1 and path 2 respectively, the detection filed can be rewritten as
\begin{equation}\label{eq0}
\hat E^{( + )} (t) = \frac{1}{{2\sqrt {2\pi } }}\int_0^\infty  {d\omega } \hat a_s (\omega )(e^{ - i\omega \tau }  + 1)e^{ - i\omega t}.
\end{equation}

The one-photon detection probability  $ P(\tau) $ is determined by
\begin{equation}\label{eq0}
  P(\tau ) = \int {dt} \left\langle {\psi \left| {\hat E^{( - )} \hat E^{( + )} } \right|\psi } \right\rangle.
\end{equation}
Consider $\hat E^{( + )} \left| \psi  \right\rangle$,
\begin{equation}\label{eq0}
\frac{1}{{2\sqrt {2\pi } }}\int_0^\infty  {d\omega } \hat a_s (\omega )(e^{ - i\omega \tau }  + 1)e^{ - i\omega t}  \times \int_0^\infty  {d\omega_s } f(\omega _s )\hat a_s^\dag  (\omega _s )\left| 0 \right\rangle  = \frac{1}{{2\sqrt {2\pi } }}\int_0^\infty  {d\omega } e^{ - i\omega t} f(\omega )(e^{ - i\omega \tau }  + 1),
\end{equation}
where the relation of  $\hat a_s (\omega )\hat a_s^\dag  (\omega _s )- \hat a_s^\dag  (\omega _s )\hat a_s (\omega ) = \delta (\omega  - \omega _s )$ is used. So,
\begin{equation}\label{eq0}
\begin{array}{lll}
 \left\langle {\psi \left| {\hat E^{( - )} \hat E^{( + )} } \right|\psi } \right\rangle & =& \frac{1}{{8\pi }}\int_0^\infty  {d\omega } e^{ - i\omega t} f(\omega )(e^{ - i\omega \tau }  + 1) \times \int_0^\infty  {d\omega ^, } e^{i\omega ^, t} f^* (\omega ^, )(e^{i\omega ^, \tau }  + 1) \\ \\
 & =& \frac{1}{{8\pi }}\int_0^\infty  \int_0^\infty  {d\omega } d\omega ^, f(\omega )f^* (\omega ^, )(e^{ - i\omega \tau }  + 1)(e^{i\omega ^, \tau }  + 1)e^{ - i(\omega  - \omega ^, )t} . \\
 \end{array}
\end{equation}
where $f^*$  is the complex conjugate of $f$ . Finally,
\begin{equation}\label{eq0}
\begin{array}{lll}
 P(\tau ) &=& \int {dt} \left\langle {\psi \left| {\hat E^{( - )} \hat E^{( + )} } \right|\psi } \right\rangle  \\\\
  &=& \frac{1}{4}\int_0^\infty \int_0^\infty  {d\omega } d\omega ^, f(\omega )f^* (\omega ^, )(e^{ - i\omega \tau }  + 1)(e^{i\omega ^, \tau }  + 1)\delta (\omega  - \omega ^, ) \\\\
  &=&\frac{1}{4}\int_0^\infty  {d\omega } f(\omega )f^* (\omega )\left| {(e^{ - i\omega \tau }  + 1)} \right|^2  \\\\
  &=& \frac{1}{2}\int_0^\infty  {d\omega } \left| {f(\omega )} \right|^2 [1 + \cos(\omega \tau )]. \\
 \end{array}
\end{equation}
In this calculation, the relation of $\delta (\omega  - \omega ^, ) = \frac{1}{{2\pi }}\int_{ - \infty }^\infty  {e^{i(\omega  - \omega ^, )t} } dt$ is used.
For a normalized $f(\omega )$, i.e.  $\int_0^\infty  {d\omega } \left| {f(\omega )} \right|^2  = 1$,
\begin{equation}\label{eq0}
P(\tau ) = \frac{1}{2}[1 + \int_0^\infty  {d\omega } \left| {f(\omega )} \right|^2 \cos(\omega \tau )].
\end{equation}
After omitting the constant component (``direct current'' component) and the coefficients, we can define the first-order correlation function as
\begin{equation}\label{eq0}
G_1 (\tau ) \equiv \int_0^\infty  {d\omega } \left| {f(\omega )} \right|^2 e^{ - i\omega \tau },
\end{equation}
where   $ P(\tau)= \frac{1}{2}[1 + Re \{ G_1(\tau) \}]$. This definition is consistent with the definition in Eq. (3.3.9) on Page 94 of Book by R. Loudon [\emph{The Quantum Theory of Light}, 3ed, \emph{Oxford}, (2000)]. The inverse Fourier transform of $G_1(\tau)$ is
\begin{equation}\label{eq0}
F_1(\omega) \equiv \left| {f(\omega )} \right|^2 = \frac{1}{2\pi}\int_0^\infty  {d\tau } G_1 (\tau )e^{i\omega \tau }
\end{equation}
This is the traditional WKT, which express power spectrum in terms of autocorrelation function by Fourier transform. Therefore, we can extract the  frequency information of the photon source from the time-domain MZ interference pattern.


\subsection*{S2: The extended Wiener-Khinchin theorem based on Hong-Ou-Mandel interference}
In Section S2, we deduce the equations for the Hong-Ou-Mandel (HOM) interference  using multi-mode theory. Based on this theory, we can construct the extended WKT (e-WKT) for differential frequency.
The setup of the HOM interference is shown in Fig.\,\ref{s1}(b).
The two-photon state from a spontaneous parametric down-conversion (SPDC) process can be described as
\begin{equation}\label{eq1}
\left| \psi  \right\rangle  = \int_0^\infty  {\int_0^\infty  {d\omega _s d\omega _i } } f(\omega _s ,\omega _i )\hat a_s^\dag  (\omega _s )\hat a_i^\dag  (\omega _i )\left| {00} \right\rangle,
\end{equation}
where $\omega$ is the angular frequency; $\hat a^\dag$ is the creation operator and the subscripts $s$ and $i$ denote the signal and idler photons from SPDC,  respectively; $f(\omega _s ,\omega _i )$ is the two-photon spectral amplitude (also called joint spectral amplitude) of the signal and idler photons.
%
%

The detection field operators of detector 1 (D1) and detector 2 (D2) are
$
\hat E_1^{( + )} (t_1 ) = \frac{1}{{\sqrt {2\pi } }}\int_0^\infty  {d\omega _1 } \hat a_1 (\omega _1 )e^{ - i\omega _1 t_1 } $ and  $\hat E_2^{( + )} (t_2 ) = \frac{1}{{\sqrt {2\pi } }}\int_0^\infty  {d\omega _2 \hat a_2 (\omega _2 )} e^{ - i\omega _2 t_2 }$, where the subscripts $1$ and $2$ denote the photons detected by D1 and D2 respectively.
The transformation rule of the 50/50 beamsplitter (BS) after a delay time $\tau$ is
$\hat a_1 (\omega _1 ) = \frac{1}{{\sqrt {2} }}[\hat a_s (\omega _1 ) + \hat a_i (\omega _1 )e^{ - i\omega _1 \tau } ]$ and $\hat a_2 (\omega _2 ) = \frac{1}{{\sqrt {2} }}[\hat a_s (\omega _2 ) - \hat a_i (\omega _2 )e^{ - i\omega _2 \tau } ]$.
So, we can rewrite the field operators as
\begin{equation}\label{eq0}
\begin{array}{lll}
\hat E_1^{( + )} (t_1 )  &=& \frac{1}{{\sqrt {4\pi } }}\int_0^\infty  {d\omega _1 } [ \hat a_s (\omega _1 )e^{ - i\omega _1 t_1 }  +  \hat a_i (\omega _1 )e^{ - i\omega _1 (t_1  + \tau )} ],\\
 \end{array}
\end{equation}
and
\begin{equation}\label{eq0}
\begin{array}{lll}
\hat E_2^{( + )} (t_2 ) &=& \frac{1}{{\sqrt {4\pi } }}\int_0^\infty  {d\omega _2 [ \hat a_s (\omega _2 )} e^{ - i\omega _2 t_2 }  - \hat a_i (\omega _2 )e^{ - i\omega _2 (t_2  + \tau )} ]. \\
 \end{array}
 \end{equation}

The two-photon detection probability $P(\tau )$ can be expressed as
\begin{equation}\label{eq0}
P(\tau ) = \int {\int {dt_1 dt_2 } } \left\langle {\psi \left| {\hat E_1^{( - )} \hat E_2^{( - )} \hat E_2^{( + )} \hat E_1^{( + )} } \right|\psi } \right\rangle.
\end{equation}
Consider $\hat E_2^{( + )} \hat E_1^{( + )} \left| \psi  \right\rangle$, only 2 out of 4 terms exist.
The first term is
\begin{equation}\label{eq0}
\begin{array}{l}
  - \frac{1}{{4\pi }}\int_0^\infty  {\int_0^\infty  {d\omega _1 d\omega _2 } } \hat a_s (\omega _1 )\hat a_i (\omega _2 )e^{ - i\omega _1 t_1 } e^{ - i\omega _2 (t_2  + \tau )} \int_0^\infty  {\int_0^\infty  {d\omega _s d\omega _i } } f(\omega _s ,\omega _i )\hat a_s^\dag  (\omega _s )\hat a_i^\dag  (\omega _i )\left| {00} \right\rangle  \\  \\
  =  - \frac{1}{{4\pi }}\int_0^\infty  {\int_0^\infty  {d\omega _1 d\omega _2 } } f(\omega _1 ,\omega _2 )e^{ - i\omega _1 t_1 } e^{ - i\omega _2 (t_2  + \tau )} \left| {00} \right\rangle.  \\
 \end{array}
\end{equation}
In the above calculation, the equations of $\hat a_s (\omega _1 )\hat a_s^\dag  (\omega _s )  \left| {0} \right\rangle = \delta (\omega _1  - \omega _s ) \left| {0} \right\rangle$  and $\hat a_i (\omega _2 )\hat a_i^\dag  (\omega _i )\left| {0} \right\rangle = \delta (\omega _2  - \omega _i ) \left| {0} \right\rangle$ are used.

The second term is
\begin{equation}\label{eq0}
\begin{array}{l}
 \frac{1}{{4\pi }}\int_0^\infty  {\int_0^\infty  {d\omega _1 } } d\omega _2 \hat a_i (\omega _1 )\hat a_s (\omega _2 )e^{ - i\omega _1 (t_1  + \tau )} e^{ - i\omega _2 t_2 } \int_0^\infty  {\int_0^\infty  {d\omega _s d\omega _i } } f(\omega _s ,\omega _i )\hat a_s^\dag  (\omega _s )\hat a_i^\dag  (\omega _i )\left| {00} \right\rangle  \\ \\
  = \frac{1}{{4\pi }}\int_0^\infty  {\int_0^\infty  {d\omega _1 } } d\omega _2 f(\omega _2 ,\omega _1 )e^{ - i\omega _1 (t_1  + \tau )} e^{ - i\omega _2 t_2 } \left| {00} \right\rangle.  \\
 \end{array}
\end{equation}
Combine these two terms:
\begin{equation}\label{eq0}
\begin{array}{ll}
 \hat E_2^{( + )} \hat E_1^{( + )} \left| \psi  \right\rangle
  &= \frac{1}{{4\pi }}\int_0^\infty  {\int_0^\infty  {d\omega _1 } } d\omega _2 e^{ - i\omega _1 t_1 } e^{ - i\omega _2 t_2 } [f(\omega _2 ,\omega _1 )e^{ - i\omega _1 \tau }  - f(\omega _1 ,\omega _2 )e^{ - i\omega _2 \tau } ] \left| {00} \right\rangle. \\
 \end{array}
\end{equation}
Then,
\begin{equation}\label{eq0}
\begin{array}{lll}
 \left\langle {\psi \left| {\hat E_1^{( - )} \hat E_2^{( - )} \hat E_2^{( + )} \hat E_1^{( + )} } \right|\psi } \right\rangle
 &= &   (\frac{1}{{4\pi }})^2 \int_0^\infty \int_0^\infty \int_0^\infty \int_0^\infty  {d\omega _1 } d\omega _2 d\omega _1^, d\omega _2^, e^{ - i(\omega _1  - \omega _1^, )t_1 } e^{ - i(\omega _2  - \omega _2^, )t_2 }\\  \\
 & & \times [ {f^* (\omega _2^, ,\omega _1^, )e^{i\omega _1^, \tau }  - f^* (\omega _1^, ,\omega _2^, )e^{i\omega _2^, \tau } } ] [ {f(\omega _2 ,\omega _1 )e^{ - i\omega _1 \tau }  - f(\omega _1 ,\omega _2 )e^{ - i\omega _2 \tau } }].  \\
 \end{array}
\end{equation}
Finally,
\begin{equation}\label{eq0}
\begin{array}{lll}
 P(\tau ) &=& \int {\int {dt_1 dt_2 } } \left\langle {\psi \left| {\hat E_1^{( - )} \hat E_2^{( - )} \hat E_2^{( + )} \hat E_1^{( + )} } \right|\psi } \right\rangle  \\ \\
   &=&\frac{1}{4}\int_0^\infty \int_0^\infty \int_0^\infty \int_0^\infty {d\omega _1 } d\omega _2 d\omega _1^, d\omega _2^, \delta (\omega _1  - \omega _1^, )\delta (\omega _2  - \omega _2^, ) \\ \\
  & & \times [ {f^* (\omega _2^, ,\omega _1^, )e^{i\omega _1^, \tau }  - f^* (\omega _1^, ,\omega _2^, )e^{i\omega _2^, \tau } }][
   {f(\omega _2 ,\omega _1 )e^{ - i\omega _1 \tau }  - f(\omega _1 ,\omega _2 )e^{ - i\omega _2 \tau } } ] \\ \\
 &=& \frac{1}{4}\int_0^\infty   \int_0^\infty   {d\omega _1 } d\omega _2 {\rm{|}}[f(\omega _1 ,\omega _2 ) - f(\omega _2 ,\omega _1 )e^{ - i(\omega _1  - \omega _2 )\tau } ]{\rm{|}}^{\rm{2}}.  \\
 \end{array}
\end{equation}
%
%
If we assume $f^* = f$, i.e. $f$ is real, we can further simplify the equation to be
\begin{equation}\label{eq0}
\begin{array}{lll}
 P(\tau)
  &=& \frac{1}{4}\int_0^\infty \int_0^\infty  {d\omega _1 } d\omega _2 [{\rm{|}}f(\omega _1 ,\omega _2 ){\rm{|}}^{\rm{2}}  + {\rm{|}}f(\omega _2 ,\omega _1 ){\rm{|}}^{\rm{2}}  - 2f(\omega _1 ,\omega _2 )f(\omega _2 ,\omega _1 )\cos(\omega _1  - \omega _2 )\tau ]. \\
 \end{array}
\end{equation}
For a normalized $f(\omega _1 ,\omega _2 )$, i.e. $\int_0^\infty \int_0^\infty  {d\omega _1 } d\omega _2 {\rm{|}}f(\omega _1 ,\omega _2 ){\rm{|}}^{\rm{2}}=1$,
\begin{equation}\label{eq0}
\begin{array}{lll}
 P(\tau)
  &=& \frac{1}{2}[ 1- \int_0^\infty \int_0^\infty  {d\omega _1 } d\omega _2     f(\omega _1 ,\omega _2 )f(\omega _2 ,\omega _1 )      \cos(\omega _1  - \omega _2 )\tau]. \\
 \end{array}
\end{equation}
If we assume $f(\omega _1 ,\omega _2 )$ has the exchange symmetry of $f(\omega _1 ,\omega _2 )= f(\omega _2 ,\omega _1 )$, we can further simplify the equation as
\begin{equation}\label{eq23}
\begin{array}{lll}
 P(\tau)
  &=& \frac{1}{2} [1-  \int_0^\infty \int_0^\infty  {d\omega _1 } d\omega _2 {\rm{|}} f(\omega _1 ,\omega _2 ) {\rm{|}}^{\rm{2}} \cos(\omega _1  - \omega _2 )\tau]. \\
 \end{array}
\end{equation}

In order to introduce less variables,  Eq.\,(\ref{eq23}) can be rewritten as
\begin{equation}\label{eq0}
\begin{array}{lll}
 P(\tau)
  &=& \frac{1}{2}[1 - \int_0^\infty \int_0^\infty  {d\omega _s } d\omega _i {\rm{|}} f(\omega _s ,\omega _i ) {\rm{|}}^{\rm{2}} \cos(\omega _s  - \omega _i )\tau]. \\
 \end{array}
\end{equation}
Next, we introduce new parameters $\omega _+= (\omega _s  + \omega _i) $ and $\omega _-=  (\omega _s  - \omega _i ) $. So, $\omega _s = \frac{1}{2} (\omega _+ + \omega _-)$ and $\omega _i = \frac{1}{2} (\omega _+ - \omega _-)$, and $P(\tau)$ can be rewritten as
\begin{equation}\label{eq0}
\begin{array}{lll}
 P(\tau)  &=& \frac{1}{2}[1 - \frac{1}{2}\int_0^\infty \int_{-\infty}^\infty  {d\omega _+ } d\omega _- {\rm{|}}f(\omega _s, \omega _i ) {\rm{|}}^{\rm{2}}   \cos( \omega _- \tau)]
  =\frac{1}{2}[1 -  \int_{-\infty}^\infty d\omega _- F_2(\omega _-)  \cos(  \omega _- \tau)], \\
 \end{array}
\end{equation}
where
\begin{equation}
F_2(\omega _-) \equiv  \frac{1}{2} \int_0^\infty  d\omega _+  {\rm{|}} f(\omega _s ,\omega _i) {\rm{|}}^{\rm{2}}
\end{equation}
is the projection of ${\rm{|}} f(\omega _s ,\omega _i ) {\rm{|}}^{\rm{2}} $ onto diagonal axis.
For a normalized $f(\omega _s ,\omega _i )$,  $F_2(\omega _-)$ is also normalized, i.e.  $\int_{-\infty}^\infty F_2(\omega _-)d\omega _- =1 $,
Note,  the following rule is used in the change of variables in the double integral,
\begin{equation}\label{eq0}
\int \int f(\omega _s ,\omega _i ) d\omega _s d\omega _i  =   \int \int f(\omega _s(\omega _ +  ,\omega _ - ), \omega _i(\omega _ +  ,\omega _ - ) ) \left| {\frac{{\partial (\omega _s ,\omega _i )}}{{\partial (\omega _ +  ,\omega _ -  )}}} \right|  d\omega _ +  d\omega _ -,
\end{equation}
which can be further simplified as
$\int \int f(\omega _s ,\omega _i )d\omega _s d\omega _i  = \frac{1}{2} \int \int f(\omega _s, \omega _i )    d\omega _ +  d\omega _ -$, because
\begin{equation}\label{eq0}
\frac{{\partial (\omega _s ,\omega _i )}}{{\partial (\omega _ +  ,\omega _ -  )}} = \left| {\begin{array}{*{20}c}
   {1/2} & {1/2}  \\
   {1/2} & { - 1/2}  \\
\end{array}} \right| =  - \frac{1}{2}
\end{equation}

So,
\begin{equation}\label{eq0}
\begin{array}{lll}
 P(\tau )
  &=&   \frac{1}{2}[1 - \int_{-\infty}^\infty d\omega _- F_2(\omega _-)  \cos( \omega _- \tau)]. \\
 \end{array}
\end{equation}

After omitting the constant component (``direct current'' component) and the coefficients, we can define the second order correlation function $ G_2(\tau)$ in the HOM interference.
\begin{equation}\label{eq0}
\begin{array}{lll}
 G_2(\tau)
  & \equiv & \int_{-\infty}^\infty d\omega _- F_2(\omega _-)  e^{-i\omega _- \tau}, \\
 \end{array}
\end{equation}
where,  $ P(\tau)= \frac{1}{2}[1 -  Re \{G_2(\tau)\}$.
The inverse Fourier transform of $G_2(\tau)$ is
\begin{equation}\label{eq0}
\begin{array}{lll}
  F_2(\omega _-)&=& \frac{1}{2\pi}\int_{-\infty}^\infty d\tau G_2(\tau) e^{i\omega _- \tau}   \\
 \end{array}
\end{equation}
This is the extended Wiener-Khinchin theorem (e-WKT) for the HOM interference, which can  provide the differential frequency information of the photon source from the time-domain HOM interference patterns.


\subsection*{S3: The extended Wiener-Khinchin theorem based on NOON-state interference}
In Section 3, we deduce the equations for the NOON-state interference  using multi-mode theory. Based on this theory, we can construct the extended Wiener-Khinchin theorem (e-WKT) for sum frequency interference.
The setup of the NOON-state interference is shown in Fig.\,\ref{s1}(c).

Assume we have the same input state as described in Section 2.
The two-photon state from a spontaneous parametric down-conversion (SPDC) process can be described as
\begin{equation}\label{eq1}
\left| \psi  \right\rangle  = \int_0^\infty  {\int_0^\infty  {d\omega _s d\omega _i } } f(\omega _s ,\omega _i )\hat a_s^\dag  (\omega _s )\hat a_i^\dag  (\omega _i )\left| {00} \right\rangle,
\end{equation}
where $\omega$ is the angular frequency; $\hat a^\dag$ is the creation operator and the subscripts $s$ and $i$ denote the signal and idler photons from SPDC,  respectively; $f(\omega _s ,\omega _i )$ is two-photon spectral amplitude (also called the join spectral amplitude) of the signal and idler photons.

The detection field operators of detector 3 (D3) and detector 4 (D4) are
$
\hat E_3^{( + )} (t_3 ) = \frac{1}{{\sqrt {2\pi } }}\int_0^\infty  {d\omega _3 } \hat a_3 (\omega _3 )e^{ - i\omega _3 t_3 } $ and  $\hat E_4^{( + )} (t_4) = \frac{1}{{\sqrt {2\pi } }}\int_0^\infty  {d\omega _4 \hat a_4 (\omega _4 )} e^{ - i\omega _4 t_4 }$, where the subscripts $3$ and $4$ denote the photons detected by D3 and D4 respectively.
The transformation rule of the second 50/50 beamsplitter (BS2) after a delay time $\tau$ is
$\hat a_3 (\omega _3 ) = \frac{1}{{\sqrt 2 }}[\hat a_1 (\omega _3 )e^{ - i\omega _3 \tau }  + \hat a_2 (\omega _3 )]$
and
$ \hat a_4 (\omega _4 ) = \frac{1}{{\sqrt 2 }}[\hat a_1 (\omega _4 )e^{ - i\omega _4 \tau }  - \hat a_2 (\omega _4 )] $.
The transformation rule of the first 50/50 beamsplitter (BS1) is
$\hat a_1 (\omega _3 ) = \frac{1}{{\sqrt 2 }}[\hat a_s (\omega _3 ) + \hat a_i (\omega _3 )]$, $\hat a_2 (\omega _3 ) = \frac{1}{{\sqrt 2 }}[\hat a_s (\omega _3 ) - \hat a_i (\omega _3 )]$,
$\hat a_1 (\omega _4 ) = \frac{1}{{\sqrt 2 }}[\hat a_s (\omega _4 ) + \hat a_i (\omega _4 )]$ and $\hat a_2 (\omega _4 ) = \frac{1}{{\sqrt 2 }}[\hat a_s (\omega _4 ) - \hat a_i (\omega _4 )]$.
So, we can rewrite the detection field operators as
\begin{equation}\label{eq0}
\begin{array}{lll}
\hat E_3^{( + )} (t_3 ) = \frac{1}{{2\sqrt {2\pi } }}\int_0^\infty  d \omega _3 [\hat a_s (\omega _3 )(e^{ - i\omega _3 \tau }  + 1)e^{ - i\omega _3 t_3 }  + \hat a_i (\omega _3 )(e^{ - i\omega _3 \tau }  - 1)e^{ - i\omega _3 t_3 } ]
 \end{array}
\end{equation}
and
\begin{equation}\label{eq0}
\begin{array}{lll}
\hat E_4^{( + )} (t_4 ) = \frac{1}{{2\sqrt {2\pi } }}\int_0^\infty  d \omega _4 [\hat a_s (\omega _4 )(e^{ - i\omega _4 \tau }  - 1)e^{ - i\omega _4 t_4 }  + \hat a_i (\omega _4 )(e^{ - i\omega _4 \tau }  + 1)e^{ - i\omega _4 t_4 } ].
 \end{array}
 \end{equation}

The two-photon detection probability $P(\tau )$ can be expressed as
\begin{equation}\label{eq0}
P(\tau )  = \int {\int {dt_3 dt_4 } } \left\langle {\psi \left| {\hat E_3^{( - )} \hat E_4^{( - )} \hat E_4^{( + )} \hat E_3^{( + )} } \right|\psi } \right\rangle.
\end{equation}
Consider $\hat E_4^{( + )} \hat E_3^{( + )} \left| \psi  \right\rangle$, only 2 out of 4 terms exist.
The first term is
\begin{equation}\label{eq0}
\begin{array}{l}
\begin{array}{l}
 \frac{1}{{8\pi }}\int_0^\infty  {\int_0^\infty  {d\omega _3 } } d\omega _4 \hat a_s (\omega _3 )(e^{ - i\omega _3 \tau }  + 1)e^{ - i\omega _3 t_3 }  \times \hat a_i (\omega _4 )(e^{ - i\omega _4 \tau }  + 1)e^{ - i\omega _4 t_4 }
 \times \int_0^\infty  {\int_0^\infty  {d\omega _s } } d\omega _i f(\omega _s ,\omega _i )\hat a_s^\dag  (\omega _s )\hat a_i^\dag  (\omega _i )\left| {00} \right\rangle  \\\\
  = \frac{1}{{8\pi }}\int_0^\infty  {\int_0^\infty  {d\omega _3 } } d\omega _4 e^{ - i\omega _3 t_3 } e^{ - i\omega _4 t_4 } f(\omega _3 ,\omega _4 )(e^{ - i\omega _3 \tau }  + 1)(e^{ - i\omega _4 \tau }  + 1) \left| {00} \right\rangle. \\
 \end{array}
 \end{array}
\end{equation}
In the above calculation, the equation  of $\hat a (\omega )\hat a ^\dag  (\omega ^, ) \left| {0} \right\rangle  = \delta (\omega   - \omega ^, ) \left| {0} \right\rangle $  is used.
The second term is
\begin{equation}\label{eq0}
\begin{array}{l}
 \frac{1}{{8\pi }}\int_0^\infty  {\int_0^\infty  {d\omega _3 } } d\omega _4 \hat a_i (\omega _3 )(e^{ - i\omega _3 \tau }  - 1)e^{ - i\omega _3 t_3 } \hat a_s (\omega _4 )(e^{ - i\omega _4 \tau }  - 1)e^{ - i\omega _3 t_3 }  \times \int_0^\infty  {\int_0^\infty  {d\omega _s } } d\omega _i f(\omega _s ,\omega _i )\hat a_s^\dag  (\omega _s )\hat a_i^\dag  (\omega _i )\left| {00} \right\rangle  \\\\
  = \frac{1}{{8\pi }}\int_0^\infty  {\int_0^\infty  {d\omega _3 } } d\omega _4 f(\omega _4 ,\omega _3 )(e^{ - i\omega _3 \tau }  - 1)(e^{ - i\omega _4 \tau }  - 1)e^{ - i\omega _3 t_3 } e^{ - i\omega _3 t_3 } \left| {00} \right\rangle.   \\
 \end{array}
\end{equation}
Combine these two terms:
\begin{equation}\label{eq0}
\begin{array}{l}
\hat E_4^{( + )} \hat E_3^{( + )} \left| \psi  \right\rangle  \\ \\
= \frac{1}{4}\int_0^\infty  {\int_0^\infty  {d\omega _3 } } d\omega _4 e^{ - i\omega _3 t_3 } e^{ - i\omega _4 t_4 } [f(\omega _3 ,\omega _4 )(e^{ - i\omega _3 \tau }  + 1)(e^{ - i\omega _4 \tau }  + 1) + f(\omega _4 ,\omega _3 )(e^{ - i\omega _3 \tau }  - 1)(e^{ - i\omega _4 \tau }  - 1)]\left| {00} \right\rangle.
 \end{array}
\end{equation}
Then,
\begin{equation}\label{eq0}
\begin{array}{lll}
\left\langle {\psi \left| {\hat E_3^{( - )} \hat E_4^{( - )} \hat E_4^{( + )} \hat E_3^{( + )} } \right|\psi } \right\rangle   \\\\
= \frac{1}{{8\pi }}\int_0^\infty  {\int_0^\infty  {d\omega _3 } } d\omega _4 e^{ - i\omega _3 t_3 } e^{ - i\omega _4 t_4 } [f(\omega _3 ,\omega _4 )(e^{ - i\omega _3 \tau }  + 1)(e^{ - i\omega _4 \tau }  + 1) + f(\omega _4 ,\omega _3 )(e^{ - i\omega _3 \tau }  - 1)(e^{ - i\omega _4 \tau }  - 1)] \\\\
\times \frac{1}{{8\pi }}\int_0^\infty  {\int_0^\infty  {d\omega _3^, } } d\omega _4^, e^{i\omega _3^, t_3 } e^{i\omega _4^, t_4 } [f^* (\omega _3^, ,\omega _4^, )(e^{i\omega _3^, \tau }  + 1)(e^{i\omega _4^, \tau }  + 1) + f^* (\omega _4^, ,\omega _3^, )(e^{i\omega _3^, \tau }  - 1)(e^{i\omega _4^, \tau }  - 1)]. \\
 \end{array}
 \end{equation}
Finally,
\begin{equation}\label{eq0}
\begin{array}{lll}
 P(\tau ) &=& \int {\int {dt_3 dt_4 } } \left\langle {\psi \left| {\hat E_3^{( - )} \hat E_4^{( - )} \hat E_4^{( + )} \hat E_3^{( + )} } \right|\psi } \right\rangle  \\ \\
  &=& \frac{1}{{16}}\int_0^\infty  {\int_0^\infty  {\int_0^\infty  {\int_0^\infty  {d\omega _3 } } } } d\omega _4 d\omega _3^, d\omega _4^, \delta (\omega _3  - \omega _3^, )\delta (\omega _4  - \omega _4^, ) \\ \\
  & \times & [f(\omega _3 ,\omega _4 )(e^{ - i\omega _3 \tau }  + 1)(e^{ - i\omega _4 \tau }  + 1) + f(\omega _4 ,\omega _3 )(e^{ - i\omega _3 \tau }  - 1)(e^{ - i\omega _4 \tau }  - 1)] \\ \\
 & \times & [f^* (\omega _3^, ,\omega _4^, )(e^{i\omega _3^, \tau }  + 1)(e^{i\omega _4^, \tau }  + 1) + f^* (\omega _4^, ,\omega _3^, )(e^{i\omega _3^, \tau }  - 1)(e^{i\omega _4^, \tau }  - 1)] \\ \\
   &=& \frac{1}{{16}}\int_0^\infty  {\int_0^\infty  {d\omega _3 } d\omega _4 } \left| {[f(\omega _3 ,\omega _4 )(e^{ - i\omega _3 \tau }  + 1)(e^{ - i\omega _4 \tau }  + 1) + f(\omega _4 ,\omega _3 )(e^{ - i\omega _3 \tau }  - 1)(e^{ - i\omega _4 \tau }  - 1)]} \right|^2.  \\
 \end{array}
\end{equation}
In the above calculation, the relation of $\delta (\omega  - \omega ^, ) = \frac{1}{{2\pi }}\int_{ - \infty }^\infty  {e^{i(\omega  - \omega ^, )t} } dt$ is used; $f^*$ is the complex conjugate of $f$.
%
%
Assuming $f$ has the symmetry of $f(\omega _3 ,\omega _4 ){\rm{ = }}f(\omega _4 ,\omega _3 )$, $ P(\tau )$ can be further simplified as
\begin{equation}\label{eq0}
P(\tau ) = \frac{1}{2}\int_0^\infty  {\int_0^\infty  {d\omega _3 } d\omega _4 } {\rm{|}}f(\omega _3 ,\omega _4 ){\rm{|}}^2 [{\rm{cos(}}\omega _3 {\rm{ + }}\omega _4 {\rm{)}}\tau {\rm{ + 1}}].
\end{equation}
For a normalized $f(\omega _3 ,\omega _4 )$, i.e.  $\int_0^\infty  {\int_0^\infty  {d\omega _3 } d\omega _4 } {\rm{|}}f(\omega _3 ,\omega _4 ){\rm{|}}^2  = 1$,
\begin{equation}\label{eq42}
P(\tau ) = \frac{1}{2}[1 + \int_0^\infty  {\int_0^\infty  {d\omega _3 } d\omega _4 } {\rm{|}}f(\omega _3 ,\omega _4 ){\rm{|}}^2 {\rm{cos(}}\omega _3 {\rm{ + }}\omega _4 {\rm{)}}\tau].
\end{equation}

In order to introduce less variables,  Eq.\,(\ref{eq42}) can be rewritten as
\begin{equation}\label{eq0}
P(\tau ) = \frac{1}{2}[1 + \int_0^\infty  {\int_0^\infty  {d\omega _s } d\omega _i } {\rm{|}}f(\omega _s ,\omega _i ){\rm{|}}^2 {\rm{cos(}}\omega _s {\rm{ + }}\omega _i {\rm{)}}\tau]
\end{equation}
Next, we use the parameters $\omega _+= (\omega _s  + \omega _i ) $ and
$\omega _- = (\omega _s  - \omega _i ) $, which are similar as  in the case of HOM interference.  So, $\omega _s = \frac{1}{2} (\omega _+ + \omega _-)$ and $\omega _i = \frac{1}{2} (\omega _+ - \omega _-)$, and $P(\tau)$ can be rewritten as
\begin{equation}\label{eq0}
\begin{array}{lll}
 P(\tau)
  = \frac{1}{2}[1 + \frac{1}{2}\int_0^\infty \int_{-\infty}^\infty  {d\omega _+ } d\omega _- \rm{|}f(\omega _s ,\omega _i ) {\rm{|}}^{\rm{2}}  \cos(  \omega _+ \tau)]
  =\frac{1}{2}[1 + \int_{-\infty}^\infty d\omega _+ F_2(\omega _+)  \cos(\omega _+ \tau)]. \\
 \end{array}
\end{equation}
where,
\begin{equation}\label{eq0}
F_2(\omega _+) \equiv  \frac{1}{2} \int_{-\infty}^\infty  d\omega _-  {\rm{|}} f(\omega _s ,\omega _i ) {\rm{|}}^{\rm{2}}
\end{equation}
is the projection of ${\rm{|}} f(\omega _s ,\omega _i ) {\rm{|}}^{\rm{2}} $ on to the anti-diagonal axis. So,
\begin{equation}\label{eq0}
\begin{array}{lll}
P(\tau )  =   \frac{1}{2}[1 +  \int_{0}^\infty d\omega _+ F_2(\omega _+)  \cos( \omega _+ \tau)]. \\
 \end{array}
\end{equation}

Omitting the constant component (``direct current'' component) and the coefficients, we can define the second-order correlation function $G_2(\tau)$ in the NOON-state interference
\begin{equation}\label{eq0}
\begin{array}{lll}
G_2(\tau)  \equiv \int_{0}^\infty d\omega _+ F_2(\omega _+)  e^{-i \omega _+ \tau}. \\
 \end{array}
\end{equation}
where,  $ P_{2}(\tau)= \frac{1}{2}[1 +  Re \{G_2(\tau)\}]$.
The inverse Fourier transform of $G_2(\tau)$ is
\begin{equation}\label{eq0}
\begin{array}{lll}
F_2(\omega _+) = \frac{1}{2\pi} \int_{0}^\infty d\tau G_2(\tau) e^{i \omega _+ \tau}. \\
 \end{array}
\end{equation}
This is the extended Wiener-Khinchin theorem for the NOON-state interference, which can  provide the sum frequency information of the photon source from the time-domain NOON-state interference patterns.


\subsection*{S4: Summary of the WKT and the e-WKT}
For simplicity in the deduction of these equations in Section S1-S3, the notifications in each section were ``local variables'',  which is valid in each section. However, for comparison, here we slightly revise the notifications and summarize these equations using ``global variables'' as follow.

    \rule{\textwidth}{0.1mm}

\textbf{The WKT}

The input state in MZ interference is
\begin{equation}\label{eq0}
\left| \psi_1  \right\rangle  = \int_{-\infty}^\infty  {d\omega  } f_1(\omega  )\hat a^\dag  (\omega  )\left| 0 \right\rangle,
\end{equation}
where $f_1(\omega  )$ is one-photon spectral amplitude. In order to keep the uniformity, here the integration range is enlarged from [0, $\infty$] to [$-\infty$, $\infty$].

The one-photon detection probability is
\begin{equation}\label{eq0}
P_1(\tau ) = \frac{1}{2}[1 + \int_{-\infty}^\infty  {d\omega } \left| {f_1(\omega )} \right|^2 \cos(\omega \tau )].
\end{equation}

The first order correlation function is
\begin{equation}\label{eq0}
G_1 (\tau ) \equiv \int_{-\infty}^\infty  {d\omega } \left| {f_1(\omega )} \right|^2 e^{ - i\omega \tau }.
\end{equation}

The connection between $P_1(\tau )$ and $G_1 (\tau )$ is
\begin{equation}\label{eq0}
 P(\tau)= \frac{1}{2}[1 + Re \{ G_1(\tau) \}].
 \end{equation}

The conventional WKT is
\begin{equation}\label{eq0}
F_1(\omega) \equiv \left| {f_1(\omega )} \right|^2 = \frac{1}{2\pi}\int_{-\infty}^\infty  {d\tau } G_1 (\tau )e^{i\omega \tau }.
\end{equation}

      \rule{\textwidth}{0.1mm}

\textbf{The e-WKT}

The input state in HOM interference and NOON-state interference is
\begin{equation}\label{eq0}
\left| \psi_2  \right\rangle  = \int_{-\infty}^\infty  {\int_{-\infty}^\infty  {d\omega _s d\omega _i } } f_2(\omega _s ,\omega _i )\hat a_s^\dag  (\omega _s )\hat a_i^\dag  (\omega _i )\left| {00} \right\rangle,
\end{equation}
where $f_2(\omega _s ,\omega _i )$ is the two-photon spectral amplitude.

The two-photon detection probability $P_2^{\pm}(\tau)$ is
\begin{equation}\label{eq0}
\begin{array}{lll}
 P_2^{\pm}(\tau)
  &=& \frac{1}{2}[1 \pm \int_{-\infty}^\infty \int_{-\infty}^\infty  {d\omega _s } d\omega _i {\rm{|}} f_2(\omega _s ,\omega _i ) {\rm{|}}^2 \cos(\omega _s  \pm \omega _i )\tau], \\
 \end{array}
\end{equation}
where   $P_2^{+}$ is for NOON-state interference, while  $P_2^{-}$ is for HOM interference.

After the transformation of variables: $\omega _{\pm}= \omega _s  \pm \omega _i $,
\begin{equation}\label{eq0}
\begin{array}{lll}
 P_2^{\pm}(\tau)
  &=& \frac{1}{2}[1 \pm \frac{1}{2}\int_{-\infty}^\infty \int_{-\infty}^\infty  {d\omega _+ } d\omega _- {\rm{|}} f_2(\omega _s ,\omega _i ) {\rm{|}}^{\rm{2}} \cos(\omega _ \pm  \tau)]. \\
 \end{array}
\end{equation}
Using the definition of the sum- or difference-frequency spectrum of the two-photon state, i.e.,
\begin{equation}\label{eq0}
 F_2^{\pm}(\omega _ \pm) \equiv \frac{1}{2}\int_{-\infty}^\infty d\omega _\mp | f_2(\omega _s ,\omega _i ) |^2  \\
\end{equation}
$P_2^{\pm}(\tau)$ can be further simplified as
\begin{equation}\label{eq0}
\begin{array}{lll}
 P_2^{\pm}(\tau)
  &=& \frac{1}{2}[1 \pm \int_{-\infty}^\infty d\omega_{\pm} F_2^{\pm}(\omega _ \pm)  \cos(\omega _ \pm  \tau)]. \\
 \end{array}
\end{equation}

Omitting the constant component (``direct current'' component) and the coefficients, we can define the second-order correlation function $G_2^{\pm}(\tau)$ as
\begin{equation}\label{eq0}
G_2^{\pm}(\tau)  \equiv  \int_{-\infty}^\infty d\omega_{\pm} F_2(\omega _{\pm})  e^{-i \omega _{\pm} \tau}. \\
\end{equation}
The connection between $P_{2}^{\pm}$ and $G_2^{\pm}$ is,
\begin{equation}\label{eq0}
 P_{2}^{\pm}(\tau)= \frac{1}{2}[1 \pm Re \{G_2^{\pm}(\tau)\}].
\end{equation}
$G_2^{+}$ is for NOON-state interference, while  $G_2^{-}$ is for HOM interference.
The inverse Fourier transform of $G_2^{\pm}(\tau)$ is
\begin{equation}\label{eq0}
F_2^{\pm}(\omega _{\pm}) = \frac{1}{2\pi} \int_{-\infty}^\infty d\tau G_2^{\pm}(\tau) e^{i \omega _{\pm} \tau}. \\
\end{equation}
This is the unified form of e-WKT.

\end{document}